\documentclass[twocolumn,english,prb,amsmath,amssymb,superscriptaddress,showpacs]{revtex4}
\usepackage[T1]{fontenc}
\usepackage[utf8x]{inputenc}
\usepackage{babel}
\usepackage{bm}
\usepackage{amsmath}
\usepackage{graphicx}
\usepackage[unicode=true,pdfusetitle,
 bookmarks=true,bookmarksnumbered=false,bookmarksopen=false,
 breaklinks=false,pdfborder={0 0 1},backref=false,colorlinks=false]
 {hyperref}
\hypersetup{
 colorlinks,citecolor=blue,linkcolor=blue,urlcolor=blue}

\makeatletter

\providecommand{\tabularnewline}{\\}

\@ifundefined{textcolor}{}
{%
 \definecolor{BLACK}{gray}{0}
 \definecolor{WHITE}{gray}{1}
 \definecolor{RED}{rgb}{1,0,0}
 \definecolor{GREEN}{rgb}{0,1,0}
 \definecolor{BLUE}{rgb}{0,0,1}
 \definecolor{CYAN}{cmyk}{1,0,0,0}
 \definecolor{MAGENTA}{cmyk}{0,1,0,0}
 \definecolor{YELLOW}{cmyk}{0,0,1,0}
}

\usepackage{bm}
\usepackage{times}
\newcommand{\ket}[1]{\lvert #1 \rangle}

\makeatother

\begin{document}
\global\long\def\MXtwo{M\! X_{2}}
\global\long\def\kp{\bm{k}\cdot\bm{p}}

\global\long\def\vr{\bm{r}}
\global\long\def\vR{\bm{R}}
\global\long\def\vk{\bm{k}}
\global\long\def\vK{\bm{K}}

\global\long\def\bktwo#1#2{\langle#1|#2\rangle}

\global\long\def\bkthree#1#2#3{\langle#1|#2|#3\rangle}

\global\long\def\ket#1{|#1\rangle}
\global\long\def\bra#1{\langle#1|}

\global\long\def\ave#1{\langle#1\rangle}

\selectlanguage{english}%

\title{Three-band tight-binding model for monolayers of group-VIB transition
metal dichalcogenides}

\date{\today}

\author{Gui-Bin Liu}

\affiliation{School of Physics, Beijing Institute of Technology, Beijing 100081,
China}

\affiliation{Department of Physics and Center of Theoretical and Computational
Physics, The University of Hong Kong, Hong Kong, China}

\author{Wen-Yu Shan}

\affiliation{Department of Physics, Carnegie Mellon University, Pittsburgh, Pennsylvania
15213, USA}

\author{Yugui Yao}

\affiliation{School of Physics, Beijing Institute of Technology, Beijing 100081,
China}

\author{Wang Yao}

\thanks{wangyao@hku.hk}

\affiliation{Department of Physics and Center of Theoretical and Computational
Physics, The University of Hong Kong, Hong Kong, China}

\author{Di Xiao}

\thanks{dixiao@cmu.edu}

\affiliation{Department of Physics, Carnegie Mellon University, Pittsburgh, Pennsylvania
15213, USA}
\begin{abstract}
We present a three-band tight-binding (TB) model for describing the
low-energy physics in monolayers of group-VIB transition metal dichalcogenides
$\MXtwo$ ($M$=Mo, W; $X$=S, Se, Te). As the conduction and valence
band edges are predominantly contributed by the $d_{z^{2}}$, $d_{xy}$,
and $d_{x^{2}-y^{2}}$ orbitals of $M$ atoms, the TB model is constructed
using these three orbitals based on the symmetries of the monolayers.
Parameters of the TB model are fitted from the first-principles energy
bands for all $\MXtwo$ monolayers. The TB model involving only the
nearest-neighbor $M$-$M$ hoppings is sufficient to capture the band-edge
properties in the $\pm K$ valleys, including the energy dispersions
as well as the Berry curvatures. The TB model involving up to the
third-nearest-neighbor $M$-$M$ hoppings can well reproduce the energy
bands in the entire Brillouin zone. Spin-orbit coupling in valence
bands is well accounted for by including the on-site spin-orbit interactions
of $M$ atoms. The conduction band also exhibits a small valley-dependent
spin splitting which has an overall sign difference between Mo$X_{2}$
and W$X_{2}$. We discuss the origins of these corrections to the
three-band model. The three-band TB model developed here is efficient
to account for low-energy physics in $\MXtwo$ monolayers, and its
simplicity can be particularly useful in the study of many-body physics
and physics of edge states.
\end{abstract}


\maketitle

\section{Introduction}

Recently, monolayers of group-VIB transition metal dichalcogenides
$\MXtwo$ ($M$ = Mo, W; $X$ = S, Se) have attracted significant
interest due to their extraordinary electronic and optical properties.
These two-dimensional semiconductors possess a direct bandgap\cite{mak_atomically_2010,splendiani_emerging_2010,Tongay_Wu_2012_12_5576__Thermally,Ross_exciton,Zeng-WX2optical}
in the visible frequency range and exhibit excellent mobility at room
temperature,\cite{radisavljevic_single_layer_2011,Lembke_Kis_2012_6_10070__Breakdown,Lin_Zhou_2012_45_345102__Mobility,Bao_Fuhrer_2013_102_42104__High,Larentis_Tutuc_2012_101_223104__Field,Fang_Javey_2012_12_3788__High}
making them promising candidates for electronic and optoelectronic
applications.\cite{Wang_Strano_2012_7_699__Electronics}

\begin{figure}
\begin{centering}
\includegraphics[width=8cm]{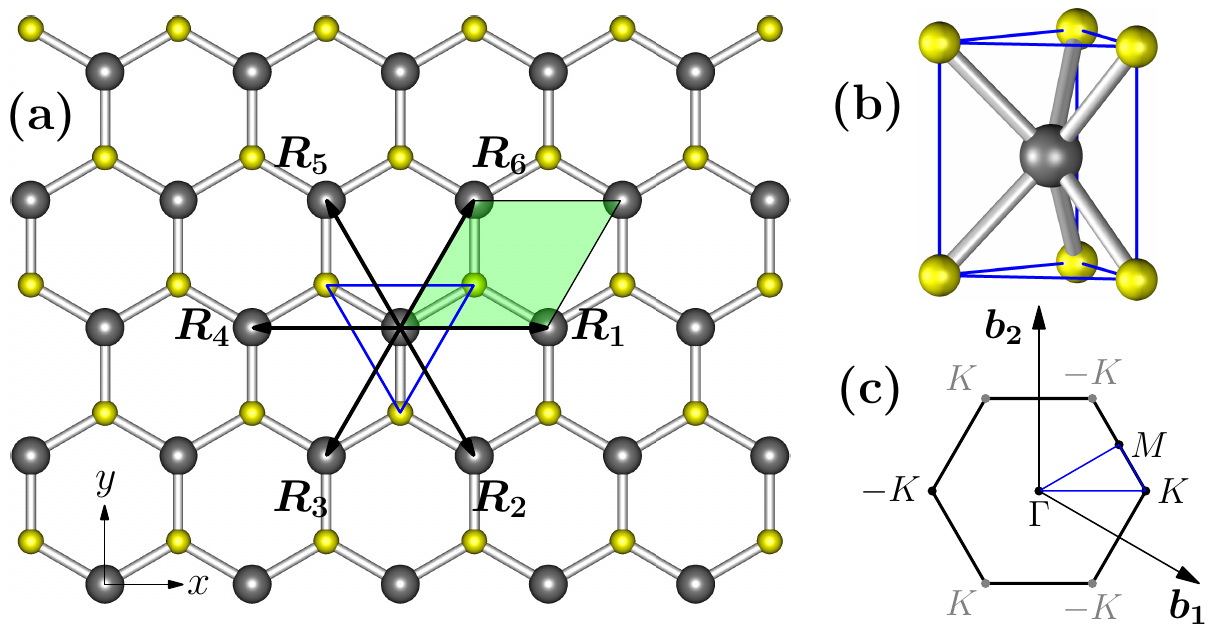}
\par\end{centering}

\caption{(color online) (a) Top view of monolayer $\MXtwo$. Big ball is $M$
and small ball is $X$. $\vR_{1}\sim\vR_{6}$ show the $M$-$M$ nearest
neighbors. The shadowed diamond region shows the 2D unit cell with
lattice constant $a$. (b) Schematic for the structure of trigonal
prismatic coordination, corresponding to the blue triangle in (a).
(c) The 2D first Brillouin zone with special $\vk$ points. $\bm{b}_{1}$
and $\bm{b}_{2}$ are the reciprocal basis vectors. The two inequivalent
valleys $K$ and $-K$ are shown in black and their equivalent counterparts
in gray. \label{fig:1}}
\end{figure}

$\MXtwo$ monolayers can be regarded as the semiconductor analog of
graphene, with both the conduction and valence band edges located
at the two corners of the first Brillouin zone (BZ), i.e. $K$ and
$-K$ points {[}Fig.\ref{fig:1}(c){]}. Thus, electrons and holes
acquire an extra valley degree of freedom, which may be used for information
encoding and processing.\cite{Xiao_Niu_2007_99_236809__Valley,Yao_Niu_2008_77_235406__Valley,Gunawan_Shayegan_2006_97_186404__Valley,Rycerz_Beenakker_2007_3_172__Valley,Behnia}
Following earlier theoretical studies,\cite{Xiao_Niu_2007_99_236809__Valley,Yao_Niu_2008_77_235406__Valley}
it was predicted that inversion symmetry breaking in monolayer $\MXtwo$
gives rise to valley dependent optical transition selection rule,
where interband transitions in $K$ and $-K$ valleys couple preferentially
to left- and right-circularly polarized light.\cite{Xiao_Yao_2012_108_196802__Coupled,cao_valley_selective_2012}
This prediction has led to the first experimental observations of
dynamical pumping of valley polarization by circularly polarized light
in monolayers of MoS$_{2}$,\cite{mak_control_2012,cao_valley_selective_2012,zeng_valley_2012}
followed by the demonstration of electric control of valley circular
dichroism in bilayer MoS$_{2}$\cite{Wu_Xu_2013_9_149__Electrical}
and valley coherence in monolayer WSe$_{2}$.\cite{Jones_intervalleycoherence}
Moreover, because of the giant spin-orbit coupling (SOC) in the material,\cite{zhu_giant_2011}
the absence of inversion symmetry also allows a strong coupling between
the spin and the valley degrees of freedom. \cite{Xiao_Yao_2012_108_196802__Coupled}
These results suggest that monolayer $\MXtwo$ could possibly be the
host for integrated spintronics and valleytronics.

In Ref. \onlinecite{Xiao_Yao_2012_108_196802__Coupled} where the
valley-spin coupled physics is first predicted in monolayer $\MXtwo$,
an effective two-band $\kp$ model is given based on symmetry considerations,
which suggests that the band edge electrons and holes can be described
as massive Dirac fermions. This $\kp$ model has also been applied
to study the transport, optical, and magnetic properties of $\MXtwo$
monolayers\cite{Li_Niu_2012_110_66803_Unconventional,Lu_Shen_2013_110_16806__Intervalley,Parhizgar_Asgari_2013_87_125401__Indirect}
and bilayers.\cite{Wu_Xu_2013_9_149__Electrical,Gong_Yao_2013___1303.3932_Magnetoelectric}
However, the $\kp$ model is only valid close to the band edge. To
obtain a more accurate description of the band structure, several
tight-binding (TB) and $\kp$ models have been recently introduced at the expense
of including more orbitals into the Hamiltonian.\cite{Rostami_Asgari_2013___1302.5901_Effective,Zahid_Guo_2013___1304.0074_generic,Kormanyos_Falko_2013___1304.4084_Monolayer,Cappelluti_Guinea_2013___1304.4831_Tight}

In this paper, we develop a minimal symmetry-based three-band TB model
using only the $M$-$d_{z^{2}}$, $d_{xy}$, and $d_{x^{2}-y^{2}}$
orbitals. We show that, by including only the nearest-neighbor (NN)
hoppings, this TB model is sufficient to capture the band-edge properties
in the $\pm K$ valleys, including the energy dispersions as well
as the Berry curvatures. By including up to the third-nearest-neighbor
(TNN) $M$-$M$ hoppings, our model can well reproduce the energy
bands in the entire BZ. All parameters in our model are determined
accurately by fitting the first-principles (FP) energy bands and results
for $X$ = Te are also shown for systematical purpose although $M$Te$_{2}$
monolayers are not realized experimentally now. SOC effects are studied
under the approximation of on-site spin-orbit interaction, which results
in a large valence-band spin splitting at the $K$ point. Besides,
for the small but finite conduction-band spin splitting at $K$ recently
noted,\cite{Cheiwchanchamnangij_Lambrecht_2012_85_205302__Quasiparticle,Kadantsev_Hawrylak_2012_152_909__Electronic,Zeng_Zhang_2012_86_241301_1209.1775_Low,Kosmider_Fernandez-Rossier_2013_87_75451__Electronic,Song_Dery_2013___1302.3627_Symmetry}
we reveal here a sign difference between Mo$X_{2}$ and W$X_{2}$,
and show that such splitting can be partly accounted for by perturbative
corrections to the three-band model. Our model provides a minimal
starting point to include various interaction effects.

This paper is organized as follows. In Sec. \ref{sec:TB}, we introduce
our three-band TB model and fitting results. In Sec. \ref{sec:SOC},
SOC effects are studied. Conclusions are given in Sec. \ref{sec:conclusions}.
In addition, an application of the TB model in zigzag nanoribbon is
demonstrated in Appendix \ref{sec:ribbon}. The relation between the
$\kp$ model in Ref. \onlinecite{Xiao_Yao_2012_108_196802__Coupled}
and this TB model is shown in Appendix \ref{sec:kp}. The FP method
is given in Appendix \ref{sec:vasp}.

\section{\label{sec:TB}The three-band TB model}

For simplicity we first introduce the spinless model and SOC will
be considered in the next Section. In the following, we first analyze
the symmetries and orbitals to determine the bases, then give the
three-band TB model involving NN $M$-$M$ hoppings, and finally introduce
up to TNN hoppings to improve the TB bands.

\subsection{Symmetries, orbitals and bases}

\begin{figure}[!tbph]
\begin{centering}
\includegraphics[width=8cm]{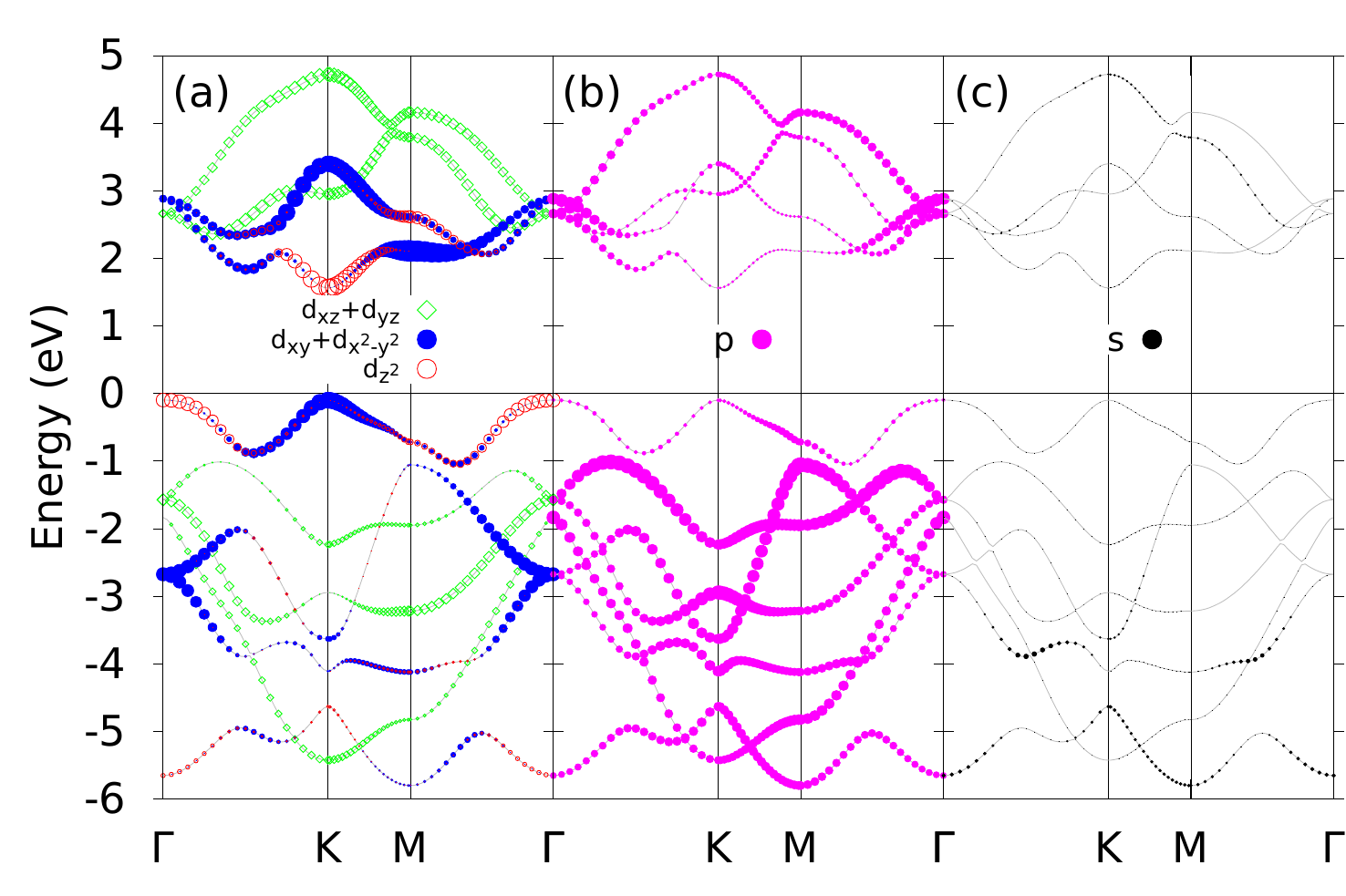}
\par\end{centering}

\caption{(color online) Orbital projected band structures for monolayer MoS$_{2}$
from FP calculations. Fermi energy is set to zero. Symbol size is
proportional to its population in corresponding state. (a) Contributions
from Mo $d$ orbitals: blue dots for $d_{xy}$ and $d_{x^{2}-y^{2}}$,
red open circles for $d_{z^{2}},$ and green open diamonds for $d_{xz}$
and $d_{yz}$. (b) Total $p$ orbitals, dominated by S atoms. (c)
Total $s$ orbitals. \label{fig:2}}
\end{figure}

Monolayer $\MXtwo$ has the $D_{3h}$ point-group symmetry and its
structure is shown in Fig. \ref{fig:1}. From early theoretical studies\cite{Bromley_Yoffe_1972_5_759__band,mattheiss_band_1973}
and recent FP investigations\cite{lebegue_electronic_2009,zhu_giant_2011,Kadantsev_Hawrylak_2012_152_909__Electronic,Ataca_Ciraci_2012_116_8983__Stable}
we know that the Bloch states of monolayer MoS$_{2}$ near the band
edges mostly consist of Mo $d$ orbitals, especially the $d_{z^{2}}$,
$d_{xy}$ and $d_{x^{2}-y^{2}}$ orbitals. Figure \ref{fig:2} clearly
shows that contributions from $s$ orbitals are negligible, those
from $p$ orbitals are very small near the band edges, and $d_{z^{2}}$,
$d_{xy}$ and $d_{x^{2}-y^{2}}$ orbitals are dominant components
for conduction and valence bands. The trigonal prismatic coordination
{[}Fig. \ref{fig:1}(b){]} splits the Mo $d$ orbitals into three
categories:\cite{mattheiss_band_1973} $A_{1}'\{d_{z^{2}}\}$, $E'\{d_{xy},d_{x^{2}-y^{2}}\}$,
and $E''\{d_{xz},d_{yz}\}$, where $A_{1}'$, $E'$, and $E''$ are
the Mulliken notations for the irreducible representations (IRs) of
point group $D_{3h}$. The reflection symmetry by the $xy$ plane,
$\hat{\sigma}_{h}$, allows hybridization only between orbitals in
$A_{1}'$ and $E'$ categories, leaving $E''$ decoupled from $A_{1}'$
and $E'$ bands {[}Fig. \ref{fig:2}(a){]}. In fact, the above analyses
are also true for all monolayers of $\MXtwo$. Therefore, it is reasonable
to construct a three-band TB model of monolayer $\MXtwo$ which can
capture the main low-energy physics by considering $d$-$d$ hoppings
using the minimal set of $M$-$d_{z^{2}}$, $d_{xy}$ and $d_{x^{2}-y^{2}}$
orbitals as bases. Obviously, using only the three $d$ orbitals and neglecting $X$-$p$ orbitals for the bases is an approximation, which will be referred as the ``three-band approximation'' hereinafter.

To conveniently describe the atomic bases by the symmetry of $D_{3h}$
point group, we denote them as $\ket{\phi_{\mu}^{j}}$ ($\mu=1,\cdots,l_{j}$)
in terms of the $\mu$-th basis belonging to the $j$-th IR:
\begin{equation}
\ket{\phi_{1}^{1}}=d_{z^{2}},\ \ \ket{\phi_{1}^{2}}=d_{xy},\ \ \ket{\phi_{2}^{2}}=d_{x^{2}-y^{2}},
\end{equation}
where $j=1$ stands for $A_{1}'$ , $j=2$ for $E'$, and $l_{j}$
for the dimension of the $j$-th IR. Then the matrix elements of the
Hamiltonian $\hat{H}$ can be easily obtained as $H_{\mu\mu'}^{jj'}(\vk)=\sum_{\vR}e^{i\vk\cdot\vR}E_{\mu\mu'}^{jj'}(\vR)$
in which
\begin{equation}
E_{\mu\mu'}^{jj'}(\vR)=\bkthree{\phi_{\mu}^{j}(\vr)}{\hat{H}}{\phi_{\mu'}^{j'}(\vr-\vR)}
\end{equation}
is the hopping integral between the atomic orbitals $\ket{\phi_{\mu}^{j}}$
at $\bm{0}$ and $\ket{\phi_{\mu'}^{j'}}$ at lattice vector $\vR$.
Given $E_{\mu\mu'}^{jj'}(\vR)$, the hopping integrals to all neighboring
sites can be generated by
\begin{equation}
E^{jj'}(\hat{g}_{n}\vR)=D^{j}(\hat{g}_{n})E^{jj'}(\vR)[D^{j'}(\hat{g}_{n})]^{\dagger},\label{eq:ERn}
\end{equation}
where $D^{j}(\hat{g}_{n})$ with dimension $l_{j}\times l_{j}$ is
the matrix of the $j$-th IR and $E^{jj'}(\vR)$ with dimension $l_{j}\times l_{j'}$
is the matrix composed of $E_{\mu\mu'}^{jj'}(\vR)$. $\hat{g}$'s
are a subset of the symmetry operations of $D_{3h}$, $\{\hat{E},\hat{C}_{3},\hat{C}_{3}^{2},\hat{\sigma}_{v},\hat{\sigma}_{v}',\hat{\sigma}_{v}''\}$,
where $\hat{E}$ is the identity operation, $\hat{C}_{3}$ is the
rotation by $2\pi/3$ around the $z$ axis, $\hat{\sigma}_{v}$ is
the reflection by the plane perpendicular to the $xy$ plane and through
the angular bisector of $\vR_{1}$ and $\vR_{6}$ in Fig. \ref{fig:1}(a),
and $\hat{\sigma}_{v}'$ and $\hat{\sigma}_{v}''$ are obtained through
rotating $\hat{\sigma}_{v}$ around $z$ axis by $2\pi/3$ and $4\pi/3$
respectively. Using the above symmetry relation, we can reduce the
parameters, i.e. the hopping integrals, to a minimal set. We emphasize
that these symmetry-based $d$-$d$ hoppings include not only the
direct $d$-$d$ interactions of $M$ atoms but also the indirect
interactions mediated by $X$-$p$ orbitals.

\subsection{Model with nearest-neighbor hoppings}

In this subsection, we introduce the three-band TB model involving
only NN $d$-$d$ hoppings, which is referred to as ``NN TB'' in
the following. After determining each Hamiltonian matrix element,
we get the three-band NN TB Hamiltonian as

\begin{equation}
H^{{\rm NN}}(\vk)=\begin{bmatrix}h_{0} & h_{1} & h_{2}\\
h_{1}^{*} & h_{11} & h_{12}\\
h_{2}^{*} & h_{12}^{*} & h_{22}
\end{bmatrix},\label{eq:HNNk}
\end{equation}
in which
\begin{equation}
h_{0}=2t_{0}(\cos2\alpha+2\cos\alpha\cos\beta)+\epsilon_{1},\label{eq:H1111}
\end{equation}
\begin{equation}
h_{1}=-2\sqrt{3}t_{2}\sin\alpha\sin\beta+2it_{1}(\sin2\alpha+\sin\alpha\cos\beta),\label{eq:H1121}
\end{equation}
\begin{equation}
h_{2}=2t_{2}(\cos2\alpha-\cos\alpha\cos\beta)+2\sqrt{3}it_{1}\cos\alpha\sin\beta,\label{eq:H1122}
\end{equation}
\begin{equation}
h_{11}=2t_{11}\cos2\alpha+(t_{11}+3t_{22})\cos\alpha\cos\beta+\epsilon_{2},\label{eq:H2121}
\end{equation}
\begin{equation}
h_{22}=2t_{22}\cos2\alpha+(3t_{11}+t_{22})\cos\alpha\cos\beta+\epsilon_{2},\label{eq:H2222}
\end{equation}

\begin{multline}
h_{12}=\sqrt{3}(t_{22}-t_{11})\sin\alpha\sin\beta\\
+4it_{12}\sin\alpha(\cos\alpha-\cos\beta),\label{eq:H2122}
\end{multline}

\begin{equation}
(\alpha,\beta)=(\frac{1}{2}k_{x}a,\frac{\sqrt{3}}{2}k_{y}a),
\end{equation}
\begin{equation}
\begin{aligned}t_{0} & =E_{11}^{11}(\vR_{1}), & t_{1} & =E_{11}^{12}(\vR_{1}), & t_{2} & =E_{12}^{12}(\vR_{1}),\\
t_{11} & =E_{11}^{22}(\vR_{1}), & t_{12} & =E_{12}^{22}(\vR_{1}), & t_{22} & =E_{22}^{22}(\vR_{1}),
\end{aligned}
\end{equation}
and $\epsilon_{j}$ is the on-site energy corresponding to the atomic
orbital $\ket{\phi_{\mu}^{j}}$. Note that, for simplicity, we have
assumed the orthogonality between each pair of different bases, therefore
the overlapping matrix of the bases is omitted and only the Hamiltonian
matrix $H^{{\rm NN}}(\vk)$ is considered. Confined by the symmetry
of the system, there are eight independent parameters in $H^{{\rm NN}}(\vk)$:
$\epsilon_{1}$, $\epsilon_{2}$, $t_{0}$, $t_{1}$, $t_{2}$, $t_{11}$,
$t_{12}$, and $t_{22}$.

\begin{table}
\caption{Band energies at the high-symmetry $\vk$ points analytically obtained
from the TB Hamiltonian Eq. \eqref{eq:HNNk}. The energies at each
$\vk$ point are in ascending order. $t_{12}>0$ is assumed.\label{tab:ene}}

\begin{ruledtabular}

\begin{centering}
\begin{tabular}{ccc}
$\Gamma=(0,0)$ & $K=(\frac{4\pi}{3a},0)$ & $M=(\frac{\pi}{a},\frac{\pi}{\sqrt{3}a})$\tabularnewline
\hline
$\epsilon_{1}+6t_{0}$ & $\epsilon_{2}-\frac{3}{2}(t_{11}+t_{22})-3\sqrt{3}t_{12}$ & $f_{1}-f_{2}$%
\footnote{$f_{1}$ and $f_{2}$ are functions independent of $t_{1}$:\\\mbox{}\hspace{1em}$f_{1}=\frac{1}{2}(\epsilon_{1}+\epsilon_{2})-t_{0}-\frac{3}{2}t_{11}+\frac{1}{2}t_{22}$,
\\\mbox{}\hspace{1em}$f_{2}=\frac{1}{2}\sqrt{(\epsilon_{1}-\epsilon_{2}-2t_{0}+3t_{11}-t_{22})^{2}+64t_{2}^{2}}$.%
}\tabularnewline
$\epsilon_{2}+3(t_{11}+t_{22})$ & $\epsilon_{1}-3t_{0}$ & $\epsilon_{2}+t_{11}-3t_{22}$\tabularnewline
 & $\epsilon_{2}-\frac{3}{2}(t_{11}+t_{22})+3\sqrt{3}t_{12}$ & $f_{1}+f_{2}$\tabularnewline
\end{tabular}
\par\end{centering}

\end{ruledtabular}
\end{table}

\begin{table*}
\caption{Fitted parameters of the three-band NN TB model based on the FP band
structures of monolayer $\MXtwo$ using both GGA and LDA. $a$ and
$z_{X\!-\! X}$ are the relaxed lattice constant and $X$-$X$ distance
in $z$ direction respectively. The energy parameters $\epsilon_{1}\sim t_{22}$
are in unit eV. \label{tab:fit}}

\begin{ruledtabular}

\begin{centering}
\begin{tabular}{ccccccccccc}
 & $a$ (\AA) & $z_{X\!-\! X}$ (\AA) & $\epsilon_{1}$ & $\epsilon_{2}$ & $t_{0}$ & $t_{1}$ & $t_{2}$ & $t_{11}$ & $t_{12}$ & $t_{22}$\tabularnewline
\hline
 & \multicolumn{10}{c}{GGA}\tabularnewline
MoS$_{2}$ & 3.190 & 3.130 & 1.046 & 2.104 & $-0.184$ & 0.401 & 0.507 & 0.218 & 0.338 & $\phantom{+}$0.057\tabularnewline
WS$_{2}$ & 3.191 & 3.144 & 1.130 & 2.275 & $-0.206$ & 0.567 & 0.536 & 0.286 & 0.384 & $-0.061$\tabularnewline
MoSe$_{2}$ & 3.326 & 3.345 & 0.919 & 2.065 & $-0.188$ & 0.317 & 0.456 & 0.211 & 0.290 & $\phantom{+}$0.130\tabularnewline
WSe$_{2}$ & 3.325 & 3.363 & 0.943 & 2.179 & $-0.207$ & 0.457 & 0.486 & 0.263 & 0.329 & $\phantom{+}$0.034\tabularnewline
MoTe$_{2}$ & 3.557 & 3.620 & 0.605 & 1.972 & $-0.169$ & 0.228 & 0.390 & 0.207 & 0.239 & $\phantom{+}$0.252\tabularnewline
WTe$_{2}$ & 3.560 & 3.632 & 0.606 & 2.102 & $-0.175$ & 0.342 & 0.410 & 0.233 & 0.270 & $\phantom{+}$0.190\tabularnewline
\hline
 & \multicolumn{10}{c}{LDA}\tabularnewline
MoS$_{2}$ & 3.129 & 3.115 & 1.238 & 2.366 & $-0.218$ & 0.444 & 0.533 & 0.250 & 0.360 & $\phantom{+}$0.047\tabularnewline
WS$_{2}$ & 3.132 & 3.126 & 1.355 & 2.569 & $-0.238$ & 0.626 & 0.557 & 0.324 & 0.405 & $-0.076$\tabularnewline
MoSe$_{2}$ & 3.254 & 3.322 & 1.001 & 2.239 & $-0.222$ & 0.350 & 0.488 & 0.244 & 0.314 & $\phantom{+}$0.129\tabularnewline
WSe$_{2}$ & 3.253 & 3.338 & 1.124 & 2.447 & $-0.242$ & 0.506 & 0.514 & 0.305 & 0.353 & $\phantom{+}$0.025\tabularnewline
MoTe$_{2}$ & 3.472 & 3.598 & 0.618 & 2.126 & $-0.202$ & 0.254 & 0.423 & 0.241 & 0.263 & $\phantom{+}$0.269\tabularnewline
WTe$_{2}$ & 3.476 & 3.611 & 0.623 & 2.251 & $-0.209$ & 0.388 & 0.442 & 0.272 & 0.295 & $\phantom{+}$0.200\tabularnewline
\end{tabular}
\par\end{centering}

\end{ruledtabular}
\end{table*}

In order to determine the eight parameters in the TB model accurately,
we fit the band structures according to the FP results. There is no
definitive strategy to fit the bands. In our case, since we are mostly
interested in the low-energy physics near the $\pm K$ points and
our analysis is entirely symmetry based, we will fit the band energies
at the high-symmetry $\vk$ points, namely $\Gamma,$ $K$, and $M$
(listed in Table \ref{tab:ene}), together with least squares fitting
according to the energies of the conduction and valence bands near
$K$.

\begin{figure*}
\centering{}\includegraphics[width=18cm]{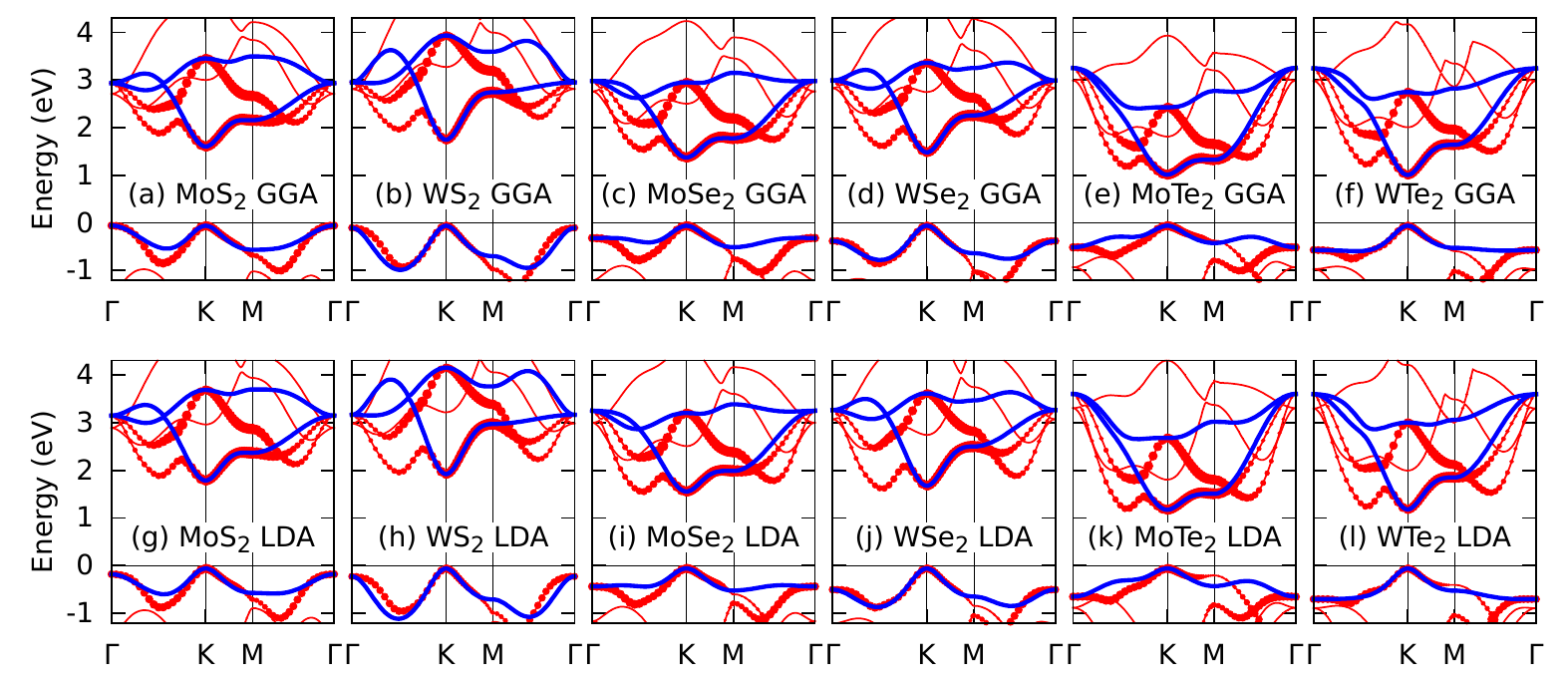}\caption{(color online) The NN TB band structures (blue or dark curves) of
$\MXtwo$ monolayers compared with the FP ones (red or gray curves
and dots). VBMs are shifted to zero. The dots show the band components
from $d_{z^{2}}$, $d_{xy}$, and $d_{x^{2}-y^{2}}$ orbitals, with
which the TB bands should compare. (a)$\sim$(f) for GGA and (g)$\sim$(l)
for LDA.\label{fig:TBband}}
\end{figure*}

By fitting the FP band structures of relaxed monolayers of $\MXtwo$
in both generalized-gradient approximation (GGA) and local-density
approximation (LDA) cases, we get the TB parameters listed in Table
\ref{tab:fit} and the corresponding band structures shown in Fig.
\ref{fig:TBband}. The FP results (lattice parameters and band structures)
obtained here are consistent with previous works.\cite{zhu_giant_2011,ding_first_2011,ataca_functionalization_2011,Ataca_Ciraci_2012_116_8983__Stable,PhysRevB.86.075454,Kang_Wu_2013_102_12111__Band}
In Fig. \ref{fig:TBband}, by comparing the TB bands with the FP bands
from $d_{z^{2}}$, $d_{xy},$ and $d_{x^{2}-y^{2}}$ orbitals, we
can see that the former agree well with the latter near the conduction-band
minimum (CBM) and valence-band maximum (VBM) at $K$ for all the $\MXtwo$
monolayers, but significantly deviate from the latter in other regions.
This is because the three-band approximation neglects the $p$ orbitals of $X$ atoms which still
have substantial contributions to the conduction bands at $\Gamma$
and valence bands at $M$ {[}Fig. \ref{fig:2}(b){]}. Nevertheless,
this simple NN TB model is sufficient to describe the physics of conduction
and valence bands in $\pm K$ valleys. In addition, a trial model
Hamiltonian of $\MXtwo$ zigzag nanoribbon based on this simple NN
TB model can give reasonable edge states (see Appendix \ref{sec:ribbon}).

We note that the band structure is very sensitive to the lattice constant:\cite{Johari_Shenoy_2012_6_5449__Tuning,Peelaers_Van_2012_86_241401__Effects,Yun_Lee_2012_85_33305__Thickness,Shi_Yakobson_2012___Quasiparticle,Horzum_Peeters_2013___1302.6635_Phonon}
in Fig. \ref{fig:TBband}(a) the valence-band energy at $\Gamma$
is higher than at $K$ by 4meV, and in Fig. \ref{fig:TBband}(i) and
\ref{fig:TBband}(j) the conduction-band energy at the dip near the
midway of $\Gamma$ and $K$ is lower than at $K$ by 5meV and 59meV
respectively. This contradicts with the observed direct bandgaps.
This is related to the different relaxed lattice constants between
GGA and LDA (GGA tends to overestimate the lattice constant whereas
LDA underestimate it, see Table \ref{tab:fit}). This, however, has
little effect on our fitting at the K point.

\subsection{Model with up to third-nearest-neighbor hoppings}

\begin{table*}
\caption{Fitted parameters (unit: eV) for the three-band TNN TB model based
on the FP bands in both GGA and LDA cases. \label{tab:TNNfit}}

\begin{ruledtabular}

\noindent \begin{centering}
\begin{tabular}{ccccccccccc}
 & $\epsilon_{1}$ & $\epsilon_{2}$ & $t_{0}$ & $t_{1}$ & $t_{2}$ & $t_{11}$ & $t_{12}$ & $t_{22}$ & $r_{0}$ & $r_{1}$\tabularnewline
 & $r_{2}$ & $r_{11}$ & $r_{12}$ & $u_{0}$ & $u_{1}$ & $u_{2}$ & $u_{11}$ & $u_{12}$ & $u_{22}$ & \tabularnewline
\hline
 &  &  &  &  & \multicolumn{2}{c}{GGA} &  &  &  & \tabularnewline
MoS$_{2}$ & 0.683  & 1.707 & -0.146 & -0.114 & 0.506 & 0.085 & 0.162 & 0.073 & 0.060 & -0.236\tabularnewline
 & 0.067 & 0.016 & 0.087 & -0.038 & 0.046 & 0.001 & 0.266 & -0.176 & -0.150 & \tabularnewline
WS$_{2}$ & 0.717  & 1.916 & -0.152 & -0.097 & 0.590 & 0.047 & 0.178 & 0.016 & 0.069 & -0.261\tabularnewline
 & 0.107 & -0.003 & 0.109 & -0.054 & 0.045 & 0.002 & 0.325 & -0.206 & -0.163 & \tabularnewline
MoSe$_{2}$ & 0.684  & 1.546 & -0.146 & -0.130 & 0.432 & 0.144 & 0.117 & 0.075 & 0.039 & -0.209\tabularnewline
 & 0.069 & 0.052 & 0.060 & -0.042 & 0.036 & 0.008 & 0.272 & -0.172 & -0.150 & \tabularnewline
WSe$_{2}$ & 0.728  & 1.655 & -0.146 & -0.124 & 0.507 & 0.117 & 0.127 & 0.015 & 0.036 & -0.234\tabularnewline
 & 0.107 & 0.044 & 0.075 & -0.061 & 0.032 & 0.007 & 0.329 & -0.202 & -0.164 & \tabularnewline
MoTe$_{2}$ & 0.588  & 1.303 & -0.226 & -0.234 & 0.036 & 0.400 & 0.098 & 0.017 & 0.003 & -0.025\tabularnewline
 & -0.169 & 0.082 & 0.051 & 0.057 & 0.103 & 0.187 & -0.045 & -0.141 & 0.087 & \tabularnewline
WTe$_{2}$ & 0.697  & 1.380 & -0.109 & -0.164 & 0.368 & 0.204 & 0.093 & 0.038 & -0.015 & -0.209\tabularnewline
 & 0.107 & 0.115 & 0.009 & -0.066 & 0.011 & -0.013 & 0.312 & -0.177 & -0.132 & \tabularnewline
\hline
 &  &  &  &  & \multicolumn{2}{c}{LDA} &  &  &  & \tabularnewline
MoS$_{2}$ & 0.820  & 1.931 & -0.176 & -0.101 & 0.531 & 0.084 & 0.169 & 0.070 & 0.070 & -0.252\tabularnewline
 & 0.084 & 0.019 & 0.093 & -0.043 & 0.047 & 0.005 & 0.304 & -0.192 & -0.162 & \tabularnewline
WS$_{2}$ & 0.905  & 2.167 & -0.175 & -0.090 & 0.611 & 0.043 & 0.181 & 0.008 & 0.075 & -0.282\tabularnewline
 & 0.127 & 0.001 & 0.114 & -0.063 & 0.047 & 0.004 & 0.374 & -0.224 & -0.177 & \tabularnewline
MoSe$_{2}$ & 0.715  & 1.687 & -0.154 & -0.134 & 0.437 & 0.124 & 0.119 & 0.072 & 0.048 & -0.248\tabularnewline
 & 0.090 & 0.066 & 0.045 & -0.067 & 0.041 & 0.005 & 0.327 & -0.194 & -0.151 & \tabularnewline
WSe$_{2}$ & 0.860  & 1.892 & -0.152 & -0.125 & 0.508 & 0.094 & 0.129 & 0.009 & 0.044 & -0.278\tabularnewline
 & 0.129 & 0.059 & 0.058 & -0.090 & 0.039 & 0.001 & 0.392 & -0.224 & -0.165 & \tabularnewline
MoTe$_{2}$ & 0.574  & 1.410 & -0.148 & -0.173 & 0.333 & 0.203 & 0.186 & 0.127 & 0.007 & -0.280\tabularnewline
 & 0.067 & 0.073 & 0.081 & -0.054 & 0.008 & 0.037 & 0.145 & -0.078 & 0.035 & \tabularnewline
WTe$_{2}$ & 0.675  & 1.489 & -0.124 & -0.159 & 0.362 & 0.196 & 0.101 & 0.044 & -0.009 & -0.250\tabularnewline
 & 0.129 & 0.131 & -0.007 & -0.086 & 0.012 & -0.020 & 0.361 & -0.193 & -0.129 & \tabularnewline
\end{tabular}
\par\end{centering}

\end{ruledtabular}
\end{table*}

\begin{figure}
\begin{centering}
\includegraphics[width=8cm]{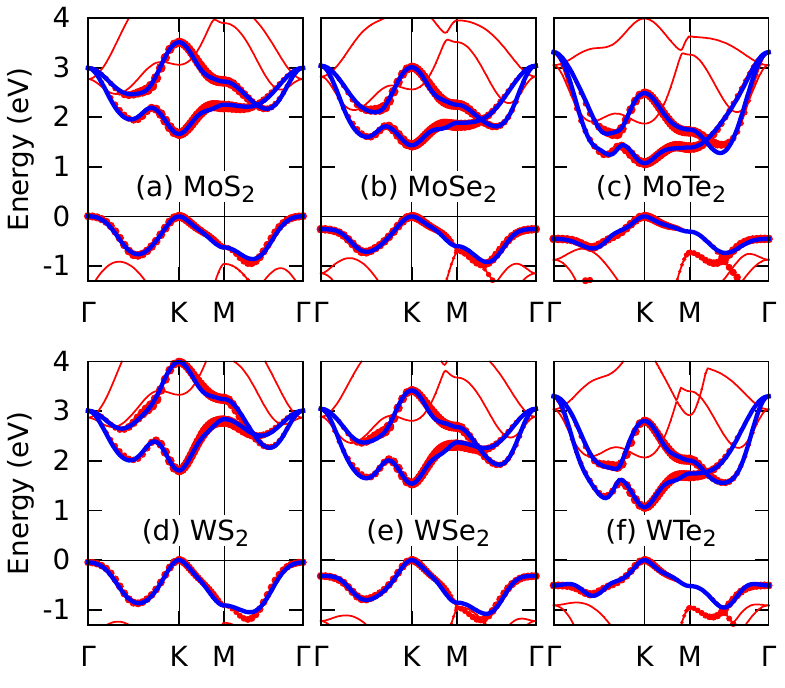}
\par\end{centering}

\caption{(color online) Energy bands from the TNN TB model (blue or dark curves)
of $\MXtwo$ monolayers compared with the FP ones in GGA case (red
or gray curves and dots). The dots show the band components from $d_{z^{2}}$,
$d_{xy}$, and $d_{x^{2}-y^{2}}$ orbitals, with which the TB bands
should compare. \label{fig:TBTNN}}
\end{figure}
\begin{figure}
\centering{}\includegraphics[width=8cm]{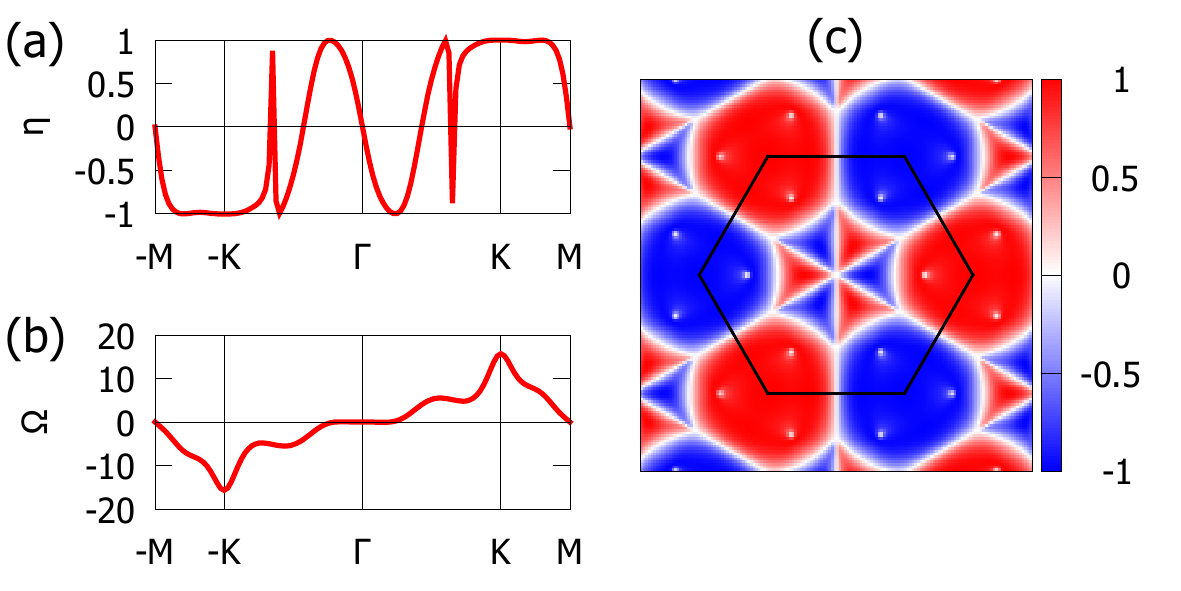}\caption{(color online) Quantities from the TNN TB for monolayer MoS$_{2}$
under GGA parameters: (a) Degree of circular polarization, $\eta(\vk)$,
and (b) Berry curvature $\Omega(\vk)$ in unit \AA$^{2}$ along $\vk$-path
$-M\rightarrow-K\rightarrow\Gamma\rightarrow K\rightarrow M$. (c)
Color map of $\eta(\vk)$ where the hexagon shows the BZ. \label{fig:eta}}
\end{figure}

In order to reproduce the energy bands in the entire BZ, we further
consider up to the TNN $M$-$M$ hoppings. By the same symmetry-based
procedure, we derive the three-band TNN model Hamiltonian $H^{\text{TNN}}(\vk)$
as
\begin{equation}
H^{\text{TNN}}(\vk)=\begin{bmatrix}V_{0} & V_{1} & V_{2}\\
V_{1}^{*} & V_{11} & V_{12}\\
V_{2}^{*} & V_{12}^{*} & V_{22}
\end{bmatrix},\label{eq:HTNNk}
\end{equation}
in which
\begin{multline}
V_{0}=\epsilon_{1}+2t_{0}(2\cos\alpha\cos\beta+\cos2\alpha)\\
+2r_{0}(2\cos3\alpha\cos\beta+\cos2\beta)\\
+2u_{0}(2\cos2\alpha\cos2\beta+\cos4\alpha),
\end{multline}
\begin{multline}
\text{Re}[V_{1}]=-2\sqrt{3}t_{2}\sin\alpha\sin\beta\\
+2(r_{1}+r_{2})\sin3\alpha\sin\beta\\
-2\sqrt{3}u_{2}\sin2\alpha\sin2\beta,
\end{multline}
\begin{multline}
\text{Im}[V_{1}]=2t_{1}\sin\alpha(2\cos\alpha+\cos\beta)\\
+2(r_{1}-r_{2})\sin3\alpha\cos\beta\\
+2u_{1}\sin2\alpha(2\cos2\alpha+\cos2\beta),
\end{multline}
\begin{multline}
\text{Re}[V_{2}]=+2t_{2}(\cos2\alpha-\cos\alpha\cos\beta)\\
-\frac{2}{\sqrt{3}}(r_{1}+r_{2})(\cos3\alpha\cos\beta-\cos2\beta)\\
+2u_{2}(\cos4\alpha-\cos2\alpha\cos2\beta),
\end{multline}
\begin{multline}
\text{Im}[V_{2}]=2\sqrt{3}t_{1}\cos\alpha\sin\beta\\
+\frac{2}{\sqrt{3}}\sin\beta(r_{1}-r_{2})(\cos3\alpha+2\cos\beta)\\
+2\sqrt{3}u_{1}\cos2\alpha\sin2\beta,
\end{multline}
\begin{multline}
V_{11}=\epsilon_{2}+(t_{11}+3t_{22})\cos\alpha\cos\beta+2t_{11}\cos2\alpha\\
+4r_{11}\cos3\alpha\cos\beta+2(r_{11}+\sqrt{3}r_{12})\cos2\beta\\
+(u_{11}+3u_{22})\cos2\alpha\cos2\beta+2u_{11}\cos4\alpha,
\end{multline}
\begin{multline}
\text{Re}[V_{12}]=\sqrt{3}(t_{22}-t_{11})\sin\alpha\sin\beta+4r_{12}\sin3\alpha\sin\beta\\
+\sqrt{3}(u_{22}-u_{11})\sin2\alpha\sin2\beta,
\end{multline}
\begin{multline}
\text{Im}[V_{12}]=4t_{12}\sin\alpha(\cos\alpha-\cos\beta)\\
+4u_{12}\sin2\alpha(\cos2\alpha-\cos2\beta),
\end{multline}
and
\begin{multline}
V_{22}=\epsilon_{2}+(3t_{11}+t_{22})\cos\alpha\cos\beta+2t_{22}\cos2\alpha\\
+2r_{11}(2\cos3\alpha\cos\beta+\cos2\beta)\\
+\frac{2}{\sqrt{3}}r_{12}(4\cos3\alpha\cos\beta-\cos2\beta)\\
+(3u_{11}+u_{22})\cos2\alpha\cos2\beta+2u_{22}\cos4\alpha.
\end{multline}
The additional parameters are defined as
\begin{equation}
\begin{aligned}r_{0} & =E_{11}^{11}(\tilde{\vR}_{1}), & r_{1} & =E_{11}^{12}(\tilde{\vR}_{1}), & r_{2} & =E_{12}^{12}(-\tilde{\vR}_{1}),\\
r_{11} & =E_{11}^{22}(\tilde{\vR}_{1}), & r_{12} & =E_{12}^{22}(\tilde{\vR}_{1})
\end{aligned}
\end{equation}
and
\begin{equation}
\begin{aligned}u_{0} & =E_{11}^{11}(2\vR_{1}), & u_{1} & =E_{11}^{12}(2\vR_{1}), & u_{2} & =E_{12}^{12}(2\vR_{1}),\\
u_{11} & =E_{11}^{22}(2\vR_{1}), & u_{12} & =E_{12}^{22}(2\vR_{1}), & u_{22} & =E_{22}^{22}(2\vR_{1}),
\end{aligned}
\end{equation}
in which $\tilde{\vR}_{1}=\vR_{1}+\vR_{2}$ is one of the next-NN
vectors and $2\vR_{1}$ is one of the TNN vectors.

The fitted parameters for $H^{{\rm TNN}}(\vk)$ are listed in Table
\ref{tab:TNNfit} and the corresponding bands are shown in Fig. \ref{fig:TBTNN}
from which we can see that the three TB bands agree well with the
FP ones contributed by $d_{z^{2}},$ $d_{xy}$ and $d_{x^{2}-y^{2}}$
orbitals in the entire BZ. The well reproduced bands mean that effective
masses can be obtained accurately by this TNN TB model. In addition,
we show the Berry curvatures calculated using this TB model in Fig.
\ref{fig:eta}(b) which shows good agreement with the result in Ref.
\onlinecite{feng_intrinsic_2012}. We note that around the $\Gamma$
point, the conduction bands with the lowest energies are made of $d_{xz}$,
$d_{yz}$ and $X$-$p$ orbitals, which cannot be captured by our
three-band model.

It should be noted that energy bands are only one aspect of physical
properties and hence not enough to capture all physics. We also calculated
the $\vk$-resolved degree of circular polarization for absorbed photons,
$\eta(\vk)$. As shown in Ref. \onlinecite{cao_valley_selective_2012},
$\eta(\vk)$ has the same sign in each region of 1/6 of the BZ around
each $K$ or $-K$ points and exhibits high degree of polarization
in most of each region. We can see that the $\eta(\vk)$ calculated
using the TB model here can give correct values in the large neighborhood
of $\pm K$, but not in the small region around $\Gamma$ {[}see Fig.
\ref{fig:eta}(a) and (c){]} due to the limitation of the three-band approximation. It can be seen from Fig. \ref{fig:TBband} and Fig. \ref{fig:eta} that the three-band approximation works well around the $\pm K$ valleys and also the valence-band $\Gamma$ point, where $d$ orbitals dominate, but not in the $k$-space region where  $X$-$p$ orbitals are important.

\section{\label{sec:SOC}Spin-orbit coupling}

\subsection{The model with SOC}

\begin{figure}
\centering{}\includegraphics[width=8cm]{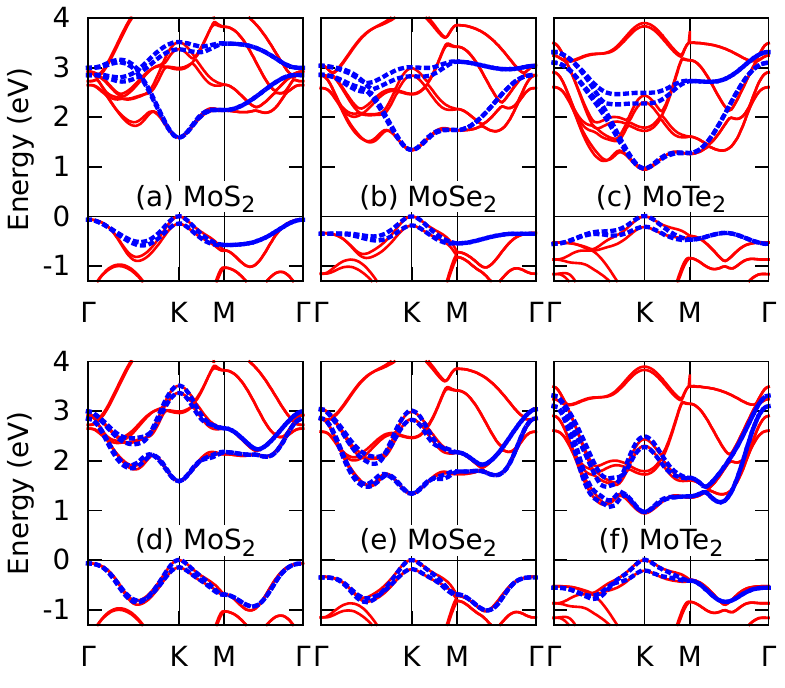}\caption{(color online) Energy bands of monolayers Mo$X_{2}$ with SOC. Thick
blue dashed curves are the TB bands: (a)\textasciitilde{}(c) from
the NN TB model and (d)\textasciitilde{}(f) from the TNN TB model.
Thin red solid curves are FP results with GGA. VBMs are shifted to
zero. \label{fig:TBsoc}}
\end{figure}

\begin{figure*}
\begin{centering}
\includegraphics[width=16cm]{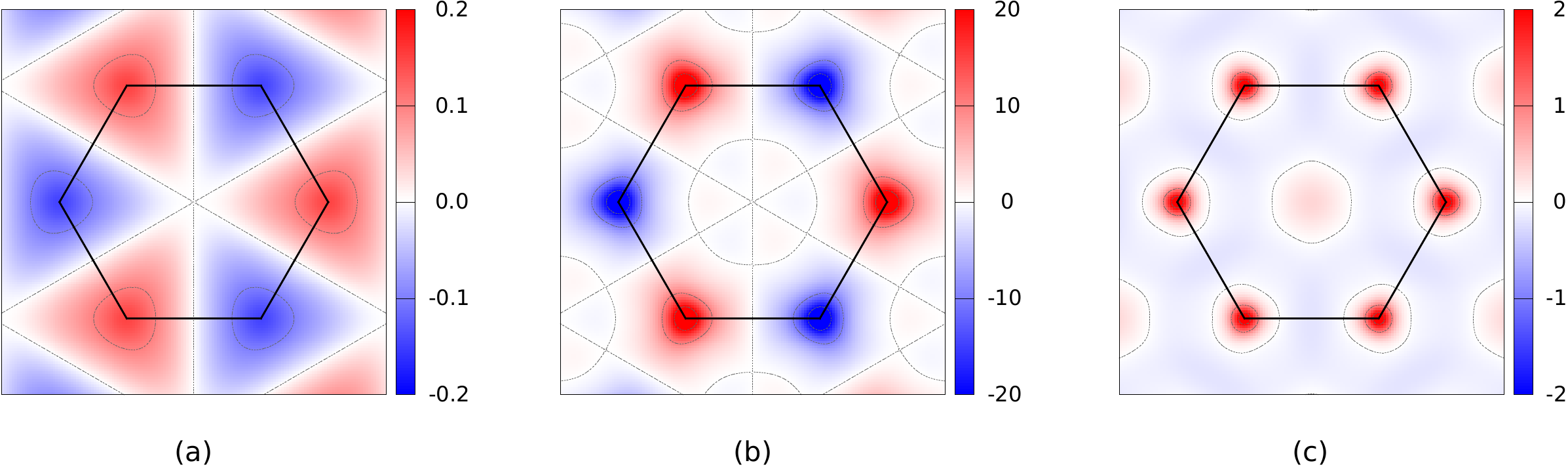}
\par\end{centering}

\caption{(color online) Contour maps in the $\vk$ space for monolayer MoS$_{2}$
from the NN TB model (using the GGA parameters): (a) the valence band
SOC splittings in unit eV, (b) the Berry curvatures and (c) the spin
Berry curvatures in unit \AA$^{2}$ . The hexagons show the BZ. The
gray thin curves are the contour lines corresponding to their tick
values on the color bars.\label{fig:BC}}
\end{figure*}
Due to the heavy transition-metal $M$ atom, its SOC can be large.
The large SOC of monolayer $\MXtwo$ is a fascinating feature which
leads to its rich physics. For simplicity, here we approximate the
SOC by considering only the on-site contribution, namely, the $\bm{L}\cdot\bm{S}$
term from $M$ atoms. Using the bases \{$\ket{d_{z^{2}},\uparrow}$,
$\ket{d_{xy},\uparrow}$, $\ket{d_{x^{2}-y^{2}},\uparrow}$, $\ket{d_{z^{2}},\downarrow}$,
$\ket{d_{xy},\downarrow}$, $\ket{d_{x^{2}-y^{2}},\downarrow}$\},
we get the SOC contribution to the Hamiltonian as
\begin{equation}
H'=\lambda\bm{L}\cdot\bm{S}=\frac{\lambda}{2}\begin{bmatrix}L_{z} & 0\\
0 & -L_{z}
\end{bmatrix},
\end{equation}
in which
\begin{equation}
L_{z}=\begin{bmatrix}0 & 0 & 0\\
0 & 0 & 2i\\
0 & -2i & 0
\end{bmatrix}
\end{equation}
is the matrix of $\hat{L}_{z}$ ($z$ component of the orbital angular
momentum) in bases of $d_{z^{2}},$ $d_{xy}$, and $d_{x^{2}-y^{2}}$,
and $\lambda$ characterizes the strength of SOC. Note that, under
the three bases, the matrix elements of $\hat{L}_{x}$ and $\hat{L}_{y}$
are all zeroes. Then we get the full TB Hamiltonian with SOC as following
\begin{eqnarray}
H_{\text{SOC}}(\vk) & = & I_{2}\otimes H_{0}(\vk)+H'\nonumber \\
 & = & \begin{bmatrix}H_{0}(\vk)+\frac{\lambda}{2}L_{z} & 0\\
0 & H_{0}(\vk)-\frac{\lambda}{2}L_{z}
\end{bmatrix},\label{eq:Hsok}
\end{eqnarray}
in which $I_{2}$ is the $2\times2$ identity matrix and $H_{0}=H^{{\rm NN}}$
or $H^{{\rm TNN}}$. The above Hamiltonian is block diagonal, which
means that the spin $z$-component is not mixed by the SOC and hence
is still a good quantum number due to the $\hat{\sigma}_{h}$ symmetry.
From Eq. \eqref{eq:Hsok} we can easily know that at $K$ point the
SOC interaction splits VBM by $\Delta_{{\rm SOC}}^{{\rm v}}=2\lambda$
and leaves CBM still degenerate (see detailed discussions in Subsection
\ref{sub:CBsoc}). The valence-band SOC (or spin) splittings are listed
in the first column of Table \ref{tab:CBsoc}. The bands from both
the NN and TNN TB Hamiltonians with SOC are shown in Fig. \ref{fig:TBsoc}
for Mo$X_{2}$. It can be seen that the NN TB bands agree well with
the FP ones only for the conduction and valence bands near the $K$
point, while the TNN TB bands agree well in the entire BZ.

Although the NN TB model is not as accurate as the TNN one, it can
still give reasonable results for low-energy physics. Taking monolayer
MoS$_{2}$ for example to test the NN TB model with SOC, we calculated
the valence band SOC splittings and the Berry curvatures and the spin
Berry curvatures, shown in Fig. \ref{fig:BC}. The valley contrasting
SOC splittings $E_{{\rm v}\uparrow}(\vk)-E_{{\rm v}\downarrow}(\vk)$
between the two spin split-off valence bands are clearly shown in
Fig. \ref{fig:BC}(a), which agrees well with the result in Ref. \onlinecite{zhu_giant_2011}.
The Berry curvatures\cite{Yao_Niu_2004_92_37204__First,Xiao_Niu_2010_82_1959__Berry}
and spin Berry curvatures\cite{Yao_Fang_2005_95_156601__Sign} are
all peaked at $\pm K$ points, and the former have opposite signs
between $K$ and $-K$ {[}Fig. \ref{fig:BC}(b){]} while the latter
have the same signs between $K$ and $-K$ {[}Fig. \ref{fig:BC}(c){]}.
These lead to valley Hall effect and spin Hall effect when an in-plane
electric field exists.\cite{Xiao_Yao_2012_108_196802__Coupled} The
TB results shown in Fig. \ref{fig:BC}(b) and (c) agree quite well
with the FP results in Ref. \onlinecite{feng_intrinsic_2012}. Therefore,
the NN TB model is sufficient to describe correctly the physics in
$\pm K$ valleys.

\subsection{\label{sub:CBsoc}The SOC splitting of conduction band}

\begin{table}
\caption{The SOC splitting of valence band at $K$ $\Delta_{{\rm SOC}}^{{\rm v}}$,
the second-order corrected SOC parameter $\lambda$, the SOC splitting
of conduction band at $K$ from the second-order perturbation theory
$\Delta_{{\rm SOC}}^{{\rm c(pt)}}$ and from FP bands $\Delta_{{\rm SOC}}^{{\rm c(FP)}}$
(GGA case), and the energy parameters in Eqs. \eqref{eq:Dvsoc} and
\eqref{eq:Dcsoc}. $E_{1,2}=E_{+1}-E_{+2}$, $E_{-1,0}=E_{-1}-E_{0}$,
and $E_{1,0}=E_{+1}-E_{0}$. All quantities are in unit eV. \label{tab:CBsoc}}

\begin{ruledtabular}

\begin{centering}
\begin{tabular}{cccccccc}
 & $\Delta_{{\rm SOC}}^{{\rm v}}$ & $\lambda$ & $\Delta_{{\rm SOC}}^{{\rm c(pt)}}$ & $\Delta_{{\rm SOC}}^{{\rm c(FP)}}$ & $E_{1,2}$ & $E_{-1,0}$ & $E_{1,0}$\tabularnewline
\hline
MoS$_{2}$ & 0.148 & 0.073 & 0.003 & $-0.003$ & 4.840 & 1.395 & 3.176\tabularnewline
WS$_{2}$ & 0.430 & 0.211 & 0.026 & $\phantom{+}$0.029 & 5.473 & 1.526 & 3.667\tabularnewline
MoSe$_{2}$ & 0.184 & 0.091 & 0.007 & $-0.021$ & 4.296 & 1.128 & 2.862\tabularnewline
WSe$_{2}$ & 0.466 & 0.228 & 0.038 & $\phantom{+}$0.036 & 4.815 & 1.267 & 3.275\tabularnewline
MoTe$_{2}$ & 0.215 & 0.107 & 0.015 & $-0.034$ & 3.991 & 0.798 & 2.918\tabularnewline
WTe$_{2}$ & 0.486 & 0.237 & 0.059 & $\phantom{+}$0.051 & 4.412 & 1.004 & 3.347\tabularnewline
\end{tabular}
\par\end{centering}

\end{ruledtabular}
\end{table}

\begin{figure}
\centering{}\includegraphics[width=8cm]{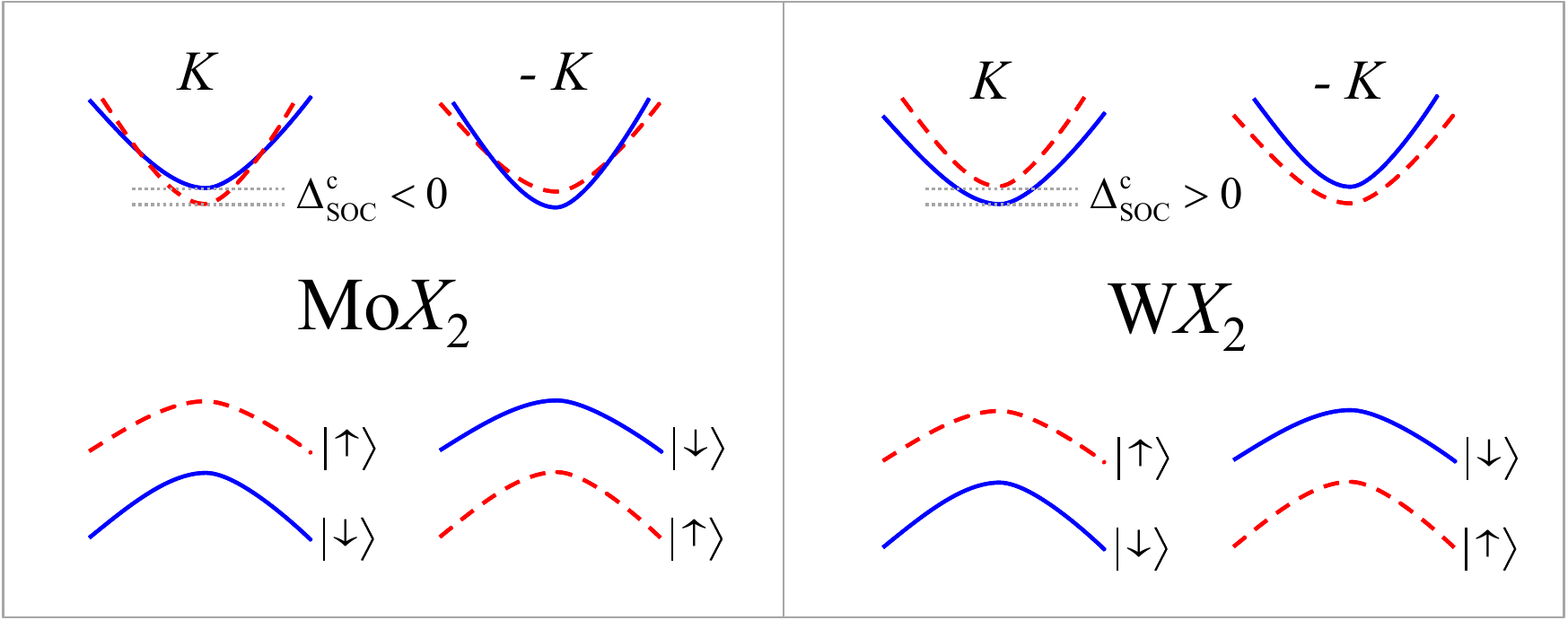}\caption{(color online) Schematic for the conduction and valence bands spin splittings in the $\pm K$ valleys for Mo$X_2$ (left panel) and W$X_2$ (right panel). Red dashed curves are spin-up states and blue solid ones spin-down states. The conduction band spin splitting has an overall sign change between Mo$X_2$ and W$X_2$. Crossings exist for the spin-split conduction bands of Mo$X_2$. \label{fig:CBSS}}
\end{figure}

To first-order of the SOC strength, the TB model for monolayer $\MXtwo$
here can only reproduce the large spin splitting of the valence band
at $K$, i.e. $\Delta_{\text{SOC}}^{{\rm v}}$, but gives no spin
splitting of the conduction band at $K$, denoted by $\Delta_{\text{SOC}}^{{\rm c}}$.
In fact, the conduction-band spin splitting (CBSS) is not zero but
a finite small value,\cite{Cheiwchanchamnangij_Lambrecht_2012_85_205302__Quasiparticle,Kadantsev_Hawrylak_2012_152_909__Electronic,Zeng_Zhang_2012_86_241301_1209.1775_Low,Kosmider_Fernandez-Rossier_2013_87_75451__Electronic,Song_Dery_2013___1302.3627_Symmetry}
and has been analyzed for MoS$_2$ by previous works. \cite{Kormanyos_Falko_2013___1304.4084_Monolayer,Ochoa2013}
Similar to the strong valley-spin coupling in the valence band,\cite{Xiao_Yao_2012_108_196802__Coupled}
the CBSS is also valley dependent due to the time-reversal symmetry
and leads to weak valley-spin coupling. Through a careful examination
of the FP results, we note here, for the first time, that the CBSSs
of Mo$X_{2}$ have opposite signs to those of W$X_{2}$, if $\Delta_{{\rm SOC}}^{{\rm c}}$
is defined as the energy difference $E_{{\rm c}\uparrow}-E_{{\rm c}\downarrow}$
at $K$ point (see Table \ref{tab:CBsoc} and Fig. \ref{fig:CBSS}). By analyzing the FP data,
we know that CBSS is induced by small contributions from $M$-$d_{xz}$,
$d_{yz}$ and $X$-$p_{x}$, $p_{y}$ orbitals. Here we will go beyond the three-band approximation and show that
a second-order perturbation correction involving $M$-$d_{xz}$ and
$d_{yz}$ orbitals can partly explain the CBSSs.

FP wavefunctions show that, at $K$ point, the Bloch waves, one below
and four above the bandgap, are dominantly composed of $d_{+2}$,
$d_{0}$, $d_{-1}$, $d_{-2}$, and $d_{+1}$ orbitals in ascending
order of energies respectively in the case without SOC,\cite{note1} in which $d_{\pm2}=\frac{1}{\sqrt{2}}(d_{x^{2}-y^{2}}\pm id_{xy})$,
$d_{0}=d_{z^{2}}$, and $d_{\pm1}=\frac{1}{\sqrt{2}}(d_{xz}\pm id_{yz})$.
Accordingly, we assume that the five Bloch states are completely composed of the aforementioned  five $d$ orbitals respectively, which is a good approximation shown by the following results.
To incorporate the contributions to CBSS from $d_{\pm1}$ (i.e. $d_{xz}$
and $d_{yz}$) orbitals, we make a second-order perturbation for the
SOC interaction $H'=\lambda\bm{L}\cdot\bm{S}$ through the L\"owdin
partitioning equation:\cite{Winkler_Winkler_2003____Spin,Xiao_Okamoto_2011_2_596__Interface}
\begin{equation}
H_{mm'}^{(2)}=\frac{1}{2}\sum_{l}H'_{ml}H'_{lm'}\Big[\frac{1}{E_{m}-E_{l}}+\frac{1}{E_{m'}-E_{l}}\Big],
\end{equation}
in which $H'_{ml}=\bkthree{d_{m}}{H'}{d_{l}}$ ($m=\pm2,0$ and $l=\pm1$)
and $E_{m}$ is the band energy at $K$ corresponding to $d_{m}$
orbital. Thus, the contributions from $d_{\pm1}$ orbitals are folded
into an effective second-order SOC interaction in bases $\{d_{+2},d_{0},d_{-2}\}\otimes\{\uparrow,\downarrow\}$
as following

\begin{multline}
H'^{(2)}={\rm diag}\{0,\,\frac{-\lambda^{2}}{E_{+1}-E_{+2}},\,\frac{-3\lambda^{2}}{2(E_{+1}-E_{0})},\\
\frac{-3\lambda^{2}}{2(E_{-1}-E_{0})},\,\frac{\lambda^{2}}{E_{-2}-E_{-1}},\,0\}.
\end{multline}
Considering the first-order SOC interaction under the same bases,
$H'^{(1)}={\rm diag}\{\lambda,-\lambda,0,0,-\lambda,\lambda\}$, finally
we get the second-order corrected splittings
\begin{equation}
\Delta_{{\rm SOC}}^{{\rm v}}=2\lambda+\frac{\lambda^{2}}{E_{+1}-E_{+2}},\label{eq:Dvsoc}
\end{equation}
\begin{equation}
\Delta_{{\rm SOC}}^{{\rm c}}=\frac{3}{2}\lambda^{2}\Big[\frac{1}{(E_{-1}-E_{0})}-\frac{1}{(E_{+1}-E_{0})}\Big].\label{eq:Dcsoc}
\end{equation}
We first get the second-order corrected $\lambda$ by solving Eq.
\eqref{eq:Dvsoc} and then put it into Eq. \eqref{eq:Dcsoc} to get
$\Delta_{{\rm SOC}}^{{\rm c}}$. The obtained {CBSSs} from perturbation
$\Delta_{{\rm SOC}}^{{\rm c(pt)}}$ are listed in Table \ref{tab:CBsoc}
and compared with the FP results $\Delta_{{\rm SOC}}^{{\rm c(FP)}}$.
The signed CBSS avoids the spurious coincidence $\Delta_{{\rm SOC}}^{{\rm c(pt)}}=|\Delta_{{\rm SOC}}^{{\rm c(FP)}}|=3$meV
for MoS$_{2}$ (see Table \ref{tab:CBsoc}). We can see that the CBSSs
determined by Eq. \eqref{eq:Dcsoc} agree very well with the FP splittings
for W$X_{2}$, but not for Mo$X_{2}$. We attribute these to the competition
of the two origins of CBSS: (i) the second-order perturbation due
to the coupling to the remote $d_{xz}$ and $d_{yz}$ orbitals; (ii)
the first-order effect from the small component of $X$-$p_{x}$ and
$p_{y}$ orbitals. Eq. \eqref{eq:Dcsoc} contains only the origin
(i) but not (ii). W atom is heavier than Mo atom, therefore the W-$d$
orbitals are the dominant contribution of the CBSSs over $X$-$p$
orbitals and thus Eq. \eqref{eq:Dcsoc} works well for W$X_{2}$.
While for Mo$X_{2}$, $X$-$p$ orbitals become non-negligible for
CBSSs relative to Mo-$d$ orbitals and Eq. \eqref{eq:Dcsoc} breaks
down for Mo$X_{2}$. More rigorous treatments involving $X$-$p$
orbitals are needed for correctly describing the CBSSs of Mo$X_{2}$,
which is out of the scope of this paper.

We also note that band crossings exist for the spin-split conduction bands of Mo$X_2$, but not for W$X_2$, as demonstrated in Fig. \ref{fig:CBSS}. The distance between the crossing and $K$ point increases from MoS$_2$ ($\sim$ 0.05 $2\pi/a$) to MoSe$_2$ ($\sim$ 0.15 $2\pi/a$), and to MoTe$_2$ ($\sim$ 0.22 $2\pi/a$). The band crossing arises from the spin dependence in the effective mass. At the $K$ point of  $\MXtwo$, the spin-down carrier has larger bandgap and thus heavier effective mass (flatter band) than the spin-up one.\cite{Xiao_Yao_2012_108_196802__Coupled} Combining the different sign of CBSS, the bands shift differently for Mo$X_2$ and W$X_2$ resulting the crossings in Mo$X_2$ but not in W$X_2$. In addition, for different Mo$X_2$, larger CBSS leads to larger distance of the crossing from the $K$ points.
Because of the trigonal warping, the distances along $K$--$\Gamma$ and $K$--$M$ directions has small difference, which is not shown in  Fig. \ref{fig:CBSS}, and crossing appears in the $K$--$M$ but not $K$--$\Gamma$ direction for MoTe$_2$ due to its relatively large CBSS.

\section{\label{sec:conclusions}conclusions}

In this paper, we have developed a minimal symmetry-based three-band
TB model for monolayers of $\MXtwo$ using only the $M$-$d_{z^{2}}$,
$d_{xy}$, and $d_{x^{2}-y^{2}}$ orbitals. When only NN $M$-$M$
hoppings are included, this TB model is sufficient to capture the
band-edge properties in the $\pm K$ valleys, including the energy
dispersions as well as the Berry curvatures. By including up to the
TNN $M$-$M$ hoppings, the model can well reproduce the energy bands
in the entire BZ. In spite of the simple NN TB model, it can describe
reasonably the edge states of zigzag $\MXtwo$ ribbon that consist
of $d_{z^{2}}$, $d_{xy}$, and $d_{x^{2}-y^{2}}$ orbitals. SOC is
introduced through the approximation of on-site $\bm{L}\cdot\bm{S}$
interactions in the heavy $M$ atoms, which lead to the giant SOC
splittings of the valence bands at $K$. In addition, we analyzed
the relatively small CBSSs at $K$ through a second-order perturbation
involving $d_{xz}$ and $d_{yz}$ orbitals, which works quite well
for W$X_{2}$ but not for Mo$X_{2}$. This is attributed to the $X$-$p$
orbitals not presented in our model. We also pointed out that the
signed CBSSs have different signs between W$X_{2}$ and Mo$X_{2}$.
The three-band TB model developed here is efficient to account for
low-energy physics in $\MXtwo$ monolayers, and its simplicity can
be particularly useful in the study of many-body physics and physics
of edge states.
\begin{acknowledgments}
The work was supported by the HKSAR Research Grant Council with Grant No. HKU706412P and the Croucher Foundation under the Croucher Innovation Award (G.B.L. and W.Y.); the National Basic Research
Program of China 973 Program with Grant No. 2013CB934500 and the Basic
Research Funds of Beijing Institute of Technology with Grant No. 20121842003
(G.B.L.); the National Basic Research Program of China 973 Program
with Grant No. 2011CBA00100, National Natural Science Foundation of
China with Grant No. 11225418 and 11174337, and Specialized Research
Fund for the Doctoral Program of Higher Education of China with Grant
No. 20121101110046 (Y.Y.); and the U.S. Department of Energy, Office
of Basic Energy Sciences, Materials Sciences and Engineering Division
(W.Y.S. and D.X.).
\end{acknowledgments}
\appendix

\section{\label{sec:ribbon}Model for zigzag nanoribbon}

\begin{figure}
\centering{}\includegraphics[width=8cm]{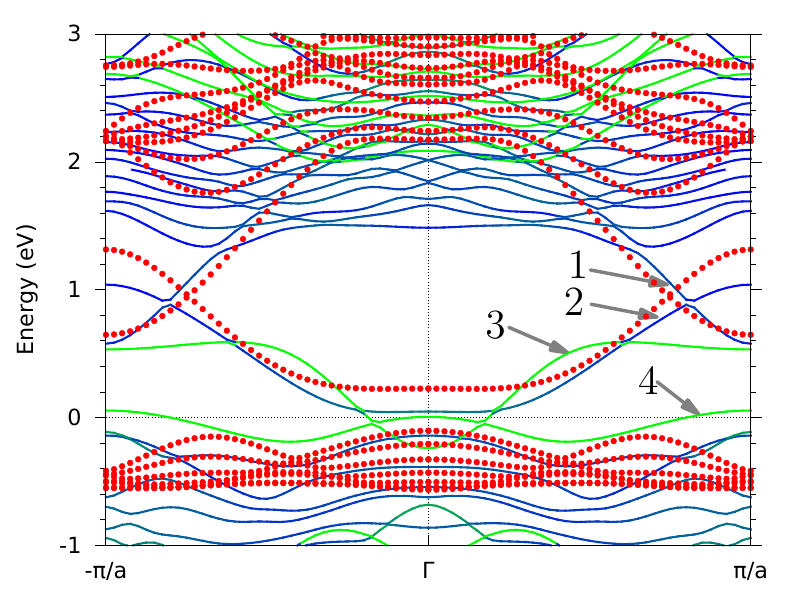}\caption{(color online) The energy bands for zigzag MoS$_{2}$ nanoribbon with
width $W=8$. Red dots are bands from the TB model using the GGA
parameters. Curves are the FP bands, in which blue color
shows the contributions from the $d_{z^{2}},$ $d_{xy}$, and $d_{x^{2}-y^{2}}$
orbitals and green color for other orbitals. For the bands labeled by $1\sim4$, see the text.\label{fig:ribbon}}
\end{figure}

In this Appendix, we apply the three-band NN TB model to study $\MXtwo$
nanoribbons. Taking a zigzag nanoribbon in $x$ direction with width
$W$ for example, there are $W$ formula units in the $y$ direction,
namely $\MXtwo\times W$, within an $x$-direction unit translational
cell. The matrix elements of Hamiltonian have three types
\begin{multline}
H_{n\gamma,n\gamma'}^{\text{ribbon }}=\delta_{\gamma\gamma'}e_{\gamma}+e^{i\vk\cdot\vR_{1}}E_{\gamma\gamma'}(\vR_{1})+e^{i\vk\cdot\vR_{4}}E_{\gamma\gamma'}(\vR_{4}),\\
(n=1,\cdots,W);
\end{multline}
\begin{multline}
H_{n\gamma,(n-1)\gamma'}^{\text{ribbon}}=e^{i\vk\cdot\vR_{2}}E_{\gamma\gamma'}(\vR_{2})+e^{i\vk\cdot\vR_{3}}E_{\gamma\gamma'}(\vR_{3}),\\
(n=2,\cdots,W);
\end{multline}
\begin{multline}
H_{n\gamma,(n+1)\gamma'}^{\text{ribbon }}=e^{i\vk\cdot\vR_{5}}E_{\gamma\gamma'}(\vR_{5})+e^{i\vk\cdot\vR_{6}}E_{\gamma\gamma'}(\vR_{6}),\\
(n=1,\cdots,W-1);
\end{multline}
in which $\gamma,\gamma'\in\{_{1}^{1},\,_{1}^{2},\,_{2}^{2}\}$, $e_{1}^{1}=\epsilon_{1}$,
and $e_{1}^{2}=e_{2}^{2}=\epsilon_{2}$. Then we can obtain the $3W\times3W$
Hamiltonian matrix for the zigzag nanoribbon as following

\begin{equation}
H^{\text{ribbon}}(k_{x})=\begin{bmatrix}h_{1}' & h_{2}'^{\dagger}\\
h_{2}' & h_{1}' & h_{2}'^{\dagger}\\
 & h_{2}' & h_{1}' & \ddots\\
 &  & \ddots & \ddots & h_{2}'^{\dagger}\\
 &  &  & h_{2}' & h_{1}'
\end{bmatrix},
\end{equation}
in which $h_{1}'\equiv H_{nn}^{\text{ribbon}}$, $h_{2}'\equiv H_{n,n-1}^{\text{ribbon}}$
and\begin{widetext}

\begin{equation}
h_{1}'=\begin{bmatrix}\epsilon_{1}+2\cos(k_{x}a)t_{0} & 2i\sin(k_{x}a)t_{1} & 2\cos(k_{x}a)t_{2}\\
-2i\sin(k_{x}a)t_{1} & \epsilon_{2}+2\cos(k_{x}a)t_{11} & 2i\sin(k_{x}a)t_{12}\\
2\cos(k_{x}a)t_{2} & -2i\sin(k_{x}a)t_{12} & \epsilon_{2}+2\cos(k_{x}a)t_{22}
\end{bmatrix},
\end{equation}

\begin{equation}
h_{2}'=\begin{bmatrix}2\cos(\frac{1}{2}k_{x}a)t_{0} & i\sin(\frac{1}{2}k_{x}a)(t_{1}-\sqrt{3}t_{2}) & -\frac{1}{2}\cos(\frac{1}{2}k_{x}a)(\sqrt{3}t_{1}+t_{2})\\
-i\sin(\frac{1}{2}k_{x}a)(t_{1}+\sqrt{3}t_{2}) & \frac{1}{2}\cos(\frac{1}{2}k_{x}a)(t_{11}+3t_{22}) & -i\sin(\frac{1}{2}k_{x}a)(\frac{\sqrt{3}}{2}t_{11}+2t_{12}-\frac{\sqrt{3}}{2}t_{22})\\
\cos(\frac{1}{2}k_{x}a)(\sqrt{3}t_{1}-t_{2}) & -i\sin(\frac{1}{2}k_{x}a)(\frac{\sqrt{3}}{2}t_{11}-2t_{12}-\frac{\sqrt{3}}{2}t_{22}) & \frac{1}{2}\cos(\frac{1}{2}k_{x}a)(3t_{11}+t_{22})
\end{bmatrix}.
\end{equation}
\end{widetext}

The energy bands of a zigzag MoS$_{2}$ nanoribbon with $W=8$ (using
the GGA parameters in Table \ref{tab:fit}) from both the TB model
and FP calculations are given in Fig. \ref{fig:ribbon}. From the
FP results, we know that the band 1 and 2 shown by arrows in Fig.
\ref{fig:ribbon} are the edge states from the $d_{z^{2}},$ $d_{xy}$,
and $d_{x^{2}-y^{2}}$ orbitals of Mo atoms at the two edges of the
ribbon, band 3 is from the Mo-$d_{yz}$ orbital at the Mo-terminated
edge, and band 4 is from the S-$p_{y}$ and $p_{z}$ orbitals at the
S-terminated edge. Due to the neglect of $d_{xz}$, $d_{yz}$ and
S-$p$ orbitals in the TB model, band 3 and 4 do not exist in the
TB bands. Nevertheless, band 1 and 2 are given by the TB model reasonably.
Therefore, the simple NN TB model for $\MXtwo$ zigzag ribbon can
give satisfactory results, if the edge states band 1 and 2 are the
focus of a study.

\section{\label{sec:kp}The two-band $k\cdot p$ model}

\begin{figure}
\centering{}\includegraphics[width=8cm]{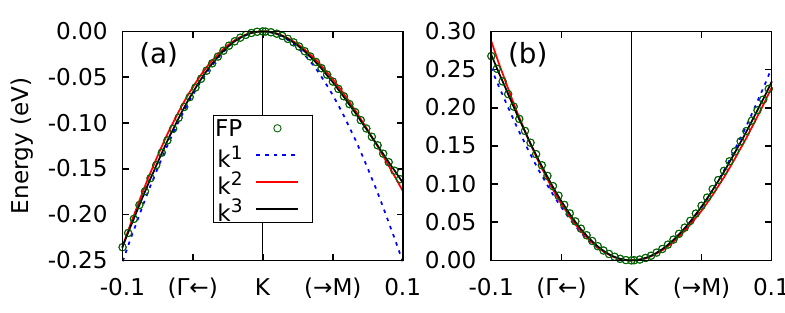}\caption{(color online) (a) Valence and (b) conduction bands in the $K$ valley
of monolayer MoS$_{2}$, within the range of 0.1$\times\frac{2\pi}{a}$
in $\Gamma$ and $M$ directions. Open circles are FP results (GGA
case). Blue dashed, red solid and black solid curves are the bands
from the two-band $\kp$ model of $H_{kp}^{(1)}$, $H_{kp}^{(2)}$
and $H_{kp}^{(3)}$ respectively. CBMs and VBMs are both shifted to
0. $a=3.190\,${\AA} and $\Delta=1.663\,$eV for all. Other fitted
parameters: $t=1.105\,$eV for $H_{kp}^{(1)}$; $t=1.059\,$eV, $\gamma_{1}=0.055\,$eV,
$\gamma_{2}=0.077\,$eV and $\gamma_{3}=-0.123\,$eV for $H_{kp}^{(2)}$;
$t=1.003\,$eV, $\gamma_{1}=0.196\,$eV, $\gamma_{2}=-0.065\,$eV,
$\gamma_{3}=-0.248\,$eV, $\gamma_{4}=0.163\,$eV, $\gamma_{5}=-0.094\,$eV
and $\gamma_{6}=-0.232\,$eV for $H_{kp}^{(3)}$. \label{fig:kp}}
\end{figure}

It is clear from Fig. \ref{fig:TBband} that the three-band NN TB
model is sufficient to describe the physics of conduction and valence
bands in the $K$ valley (also true for $-K$ valley due to the time
reversal symmetry). Thus we can expand Eq. \eqref{eq:HNNk} in the
$\pm K$ valleys to any order required and then reduce it to a two-band
$\kp$ model in the L\"owdin partitioning method.\cite{Lowdin_Lowdin_1951_19_1396__Note,Kohn_canonical_transformation,Winkler_Winkler_2003____Spin}
Using $\ket{\psi_{{\rm c}}^{\tau}}=\ket{d_{z^{2}}}$ and $\ket{\psi_{{\rm v}}^{\tau}}=\frac{1}{\sqrt{2}}(\ket{d_{x^{2}-y^{2}}}+i\tau\ket{d_{xy}})$
($\tau=\pm$ is the valley index) as bases, the obtained two-band
$\kp$ model up to the third order in $\vk$ (relative to $\tau K$)
are

\begin{equation}
H_{kp}^{(1)}(\vk;\tau)=\begin{bmatrix}\Delta/2 & at(\tau k_{x}-ik_{y})\\
at(\tau k_{x}+ik_{y}) & -\Delta/2
\end{bmatrix},\label{eq:Hkp1}
\end{equation}
\begin{multline}
H_{kp}^{(2)}(\vk;\tau)=H_{kp}^{(1)}(\vk;\tau)+\\
a^{2}\begin{bmatrix}\gamma_{1}k^{2} & \gamma_{3}(\tau k_{x}+ik_{y})^{2}\\
\gamma_{3}(\tau k_{x}-ik_{y})^{2} & \gamma_{2}k^{2}
\end{bmatrix},\label{eq:Hkp2}
\end{multline}
\begin{multline}
H_{kp}^{(3)}(\vk;\tau)=H_{kp}^{(2)}(\vk;\tau)+\\
a^{3}\begin{bmatrix}\gamma_{4}\tau k_{x}(k_{x}^{2}-3k_{y}^{2}) & \gamma_{6}k^{2}(\tau k_{x}-ik_{y})\\
\gamma_{6}k^{2}(\tau k_{x}+ik_{y}) & \gamma_{5}\tau k_{x}(k_{x}^{2}-3k_{y}^{2})
\end{bmatrix},\label{eq:Hkp3}
\end{multline}
in which $\Delta$ is the bandgap at $K$, $t$ and $\gamma_{1}\sim\gamma_{6}$
are energy parameters, and $k^{2}=k_{x}^{2}+k_{y}^{2}$. Eq. \eqref{eq:Hkp1}
is the massive Dirac Hamiltonian given in Ref. \onlinecite{Xiao_Yao_2012_108_196802__Coupled}
which was derived just this way, and Eqs. \eqref{eq:Hkp2} and \eqref{eq:Hkp3}
are consistent with previous works.\cite{Rostami_Asgari_2013___1302.5901_Effective,Kormanyos_Falko_2013___1304.4084_Monolayer}
In Fig. \ref{fig:kp}, the bands of monolayer MoS$_{2}$ from $H_{kp}^{(1)}$
capture the main physics in the valley but neglect the details such
as the  anisotropic dispersion (the trigonal warping) and the
 electron-hole asymmetry, the bands from $H_{kp}^{(2)}$ recover
the aforementioned missing details, and the bands from $H_{kp}^{(3)}$
agree with the FP bands perfectly.

When SOC is considered to the first order, Eq. \eqref{eq:Hsok} is
still valid and we can get
\begin{equation}
H_{kpso}^{(n)}(\vk;\tau,s)=H_{kp}^{(n)}(\vk;\tau)+\begin{bmatrix}0 & 0\\
0 & \tau s\lambda
\end{bmatrix},\label{eq:Hkp2soc}
\end{equation}
where $s=\pm1$ is the spin index ($+1$ for $\uparrow$ and $-1$
for $\downarrow$) since spin is a good quantum number. The $\tau s\lambda$
term in Eq. \eqref{eq:Hkp2soc} appears in the form of the product
of the valley index $\tau$, the spin index $s$, and the SOC parameter
$\lambda$, which implies the rich physics due to the SOC induced
coupling of valley and spin described in Ref. \onlinecite{Xiao_Yao_2012_108_196802__Coupled}.

\section{\label{sec:vasp}FP band structure calculations}

The FP band structures used for fitting the parameters were calculated
by the VASP package\cite{vasp1,vasp2} using the projector-augmented
wave (PAW) method.\cite{PAW,vaspPAW} Exchange-correlation functionals
of both GGA\cite{PBE96} and LDA\cite{Ceperley_Alder_1980_45_566__Ground,Perdew_Zunger_1981_23_5048__Self}
were used to give comparable results. The energy cutoff of plane wave
basis was set to 400 eV and the convergence criterion $10^{-6}$ eV.
A gamma-centered $\vk$-mesh of $10\times10\times1$ was used and
layer separation was greater than 15 \AA. For all monolayers of $\MXtwo$,
lattice constants were optimized and atomic positions were relaxed
until the force on each atom was less than 0.005 eV/\AA.


\begin{thebibliography}{65}
\bibitem{note1} For WTe$_2$ monolayer, the band with $d_{+1}$ orbital above the bandgap is the 5th band above the bandgap, while it is the 4th band above the bandgap for other $\MXtwo$ monolayers.
\end{thebibliography}

%

\end{document}